\documentclass[fdp,fleqn]{w-art}
\usepackage{times}
\usepackage{w-thm}
\def\AEF{A.E. Faraggi}
\def\JHEP#1#2#3{\emph{JHEP}\/ {\bf #1}, #3 (#2)}
\def\NPB#1#2#3{\emph{ Nucl.\ Phys.}\/ B {\bf #1}, #3 (#2)}
\def\PLB#1#2#3{\emph{ Phys.\ Lett.}\/ B {\bf #1}, #3 (#2)}
\def\PRD#1#2#3{\emph{ Phys.\ Rev.}\/ D {\bf #1}, #3 (#2)}
\def\PRL#1#2#3{\emph{ Phys.\ Rev.\ Lett.}\/ {\bf #1}, #3 (#2)}

\def\IJMP#1#2#3{\emph{ Int.\ J.\ Mod.\ Phys.}\/ A {\bf #1}, #3 (#2)}

\def\EPJC#1#2#3{\emph{Eur.\ Phys.\ Jour.}\/ C {\bf #1}, #3 (#2)}

\usepackage[]{graphicx}
\begin{document}
\DOIsuffix{theDOIsuffix}
\Volume{55}
\Issue{1}
\Month{01}
\Year{2007}
\pagespan{1}{}
\keywords{String phenomenology, spinor--vector duality}



\title[Equivalence of string vacua]{On the Equivalence of String Vacua}


\author[A.E. faraggi]{Alon E. Faraggi\inst{1}%
  \footnote{Corresponding author\quad E-mail:~\textsf{alon.faraggi@liv.ac.uk}, 
            Phone: +00\,44\,0151\,7943774, 
            Fax: +00\,44\,0151\,7943784}}
\address[\inst{1}]{Department of
Mathematical Sciences, University of Liverpool, Liverpool L69 7ZL, UK}
\begin{abstract}
The heterotic--string models in the free fermionic formulation
are among the most realistic string vacua constructed to date,
which motivates their detailed exploration. Classification
of free fermionic heterotic--string vacua revealed a duality 
under exchange of spinor and vector representations of the 
$SO(10)$ GUT symmetry. The spinor--vector duality was 
subsequently demonstrated in a $Z_2\times Z_2\times Z_2$ orbifold,
in which the map is realised as exchange of discrete torsions.
Analysis of the orbifold partition function shows that the 
duality map preserves the number of massless twisted states. 
While the dual vacua are distinct from the point of view
of the low energy field theory it is suggested that they are 
equivalent from the string point of view and may be connected
by continuous and discrete transformations. 

\end{abstract}
\maketitle                   






String theory provides a unique framework to study 
the unification of the gauge and gravitational interactions.
Progress in this endeavour mandates the development
of phenomenological models as well as improved insight
into the mathematical structures that underlie the theory.
The formal understanding of string theory is still very
much in its infancy, as is, evidently, of any of its 
phenomenological implications. A variety of world--sheet and
target--space tools are used to advance these investigations. 
Analysis of the effective field theory limit of
string vacua uses in particular target--space techniques,
whereas world--sheet methods provides access to
the massive string spectrum. A particular area of
ignorance is the relation between the string vacua
and their low energy effective field theory.
This relation is fairly well understood only in
very special simple cases. Specifically, the cases
with (2,2) world--sheet super--conformal field theory
correspond to compactifications on Calabi--Yau 
manifolds. 
However, in the more generic
case, with only (2,0) world--sheet supersymmetry,
the relation is, in general, obscure.
This issue is vital for developing the methodology
to confront string theory with observational data. 

The relationship between the string theory vacua and their
low energy effective field theory limit also raises the question
in regard to the enumeration of solutions. It is well known
that string theory gives rise to a multitude of vacua,
both in the world--sheet approach and in the effective field 
theory description. However, the pictures are not identical
in the two regimes. While two vacuum states in the effective 
field theory limit may correspond to distinct physical
spectra, the string solution allows access to the massive
sector, and the two vacua may be connected under interchange 
of massless and massive states. This notion is of course not new
and has appeared in the form of perturbative and non--perturbative
dualities in the past, as well as in the form of 
topology changing transitions. 
These various types of dualities are key to the deeper 
understanding of string theory, in particular, and of 
quantum gravity, in general. It was also suggested that
T--duality may be regarded as phase space duality  
in compact space. A formalism that promotes phase-space
duality to a level of fundamental principle 
was developed in ref. \cite{faraggimatone}
and used as a starting point for a derivation 
of quantum mechanics from an equivalence postulate 
\cite{faraggimatone}.

Over the past few years a new duality symmetry in the space
of heterotic--string vacua was discovered, under the exchange of 
the total number of spinor plus anti--spinor with the 
total number of vector representations of a $SO(10)$ Grand Unified 
Theory (GUT) group. The spinor--vector duality was first observed 
in classification of fermionic $Z_2\times Z_2$ heterotic--string
orbifolds. The free fermionic formalism \cite{abk,klt}
was used since the late eighties
to construct some of the most realistic string models to date  
\cite{fsu5,fny,alr,nahe,eu,top,cslm,cfn,cfs,su421,fmt,cfmt}.
These models provide a
concrete framework to study many of the issues that pertain to the
phenomenology of the Standard Model
and grand unification. A few highlights of these studies
are listed below:

\begin{itemize}
\item[{$\bullet$}] { Top quark mass $\sim$ 175--180GeV  \cite{top}}, 
                   { Generation mass hierarchy \cite{fmm}} \& 
                   { CKM mixing  \cite{ckm,fh}} 
\item[{$\bullet$}] { Stringy seesaw mechanism \cite{seesaw,fh2,seesawII}},
                   { Gauge coupling unification \cite{gcu,df}} \&
                   { Proton stability \cite{ps}}

\item[{$\bullet$}] { Squark degeneracy \cite{fp2,fv}} \&
                   { Moduli fixing \cite{modulifixing}}
\item[{$\bullet$}] { Minimal Standard Heterotic String Model (MSHSM) 
                                                                \cite{cfn}}
\item[{$\bullet$}] { Classification \cite{nooij, classification}}, 
                   { spinor--vector duality \cite{spduality,tristan}} \&
                   { Exophobia \cite{exophobic}}
\end{itemize}

These results accentuate the need for deeper understanding of 
this class of models. In the free fermionic formalism the modular invariance 
constraints are solved in terms of the
transformation properties of the world--sheet fermions on the string
world--sheet, and are encoded in sets of basis vectors and
one--loop GSO projection coefficients \cite{abk,klt}.
The free fermionic formalism is adaptable to a computerised 
classification of $Z_2\times Z_2$ orbifolds with symmetric
shifts \cite{nooij, classification}.
It corresponds to using
a free bosonic formalism in which the radii of the internal dimensions 
are fixed at a special point in the compact space.
Deformation from the special point in the moduli space are parametrized 
in terms of world--sheet Thirring interactions among the world--sheet 
fermions \cite{klt}. The equivalence of bosons and fermions in two dimensions 
entails that a model constructed
using the fermionic approach corresponds to a model
constructed using the bosonic approach in which the
target--space is compactified on a six dimensional internal manifold.

The free fermionic models have also been instrumental in recent years to
unravel a new duality symmetry under the exchange of spinor and vector 
representations of the GUT group \cite{spduality, tristan}.
This has been achieved by generating large number of vacua and
observing the symmetry over the entire class. An algebraic 
proof of the spinor--vector duality map was given in terms of
the GGSO projection coefficients \cite{spduality},
as well as an operational 
representation in terms of free phases in the one--loop
partition function \cite{tristan}. 

Further insight into the properties of the free fermionic models,
in particular with respect to moduli dynamics,
can be achieved by obtaining orbifold representations
of these models.
In the orbifold models one typically starts with the
$E_8\times E_8$ heterotic--string compactified to
four dimensions and subsequently breaks one $E_8$ gauge factor
by the orbifold twistings and discrete Wilson lines.
In contrast the quasi--realistic free fermionic models
start with  $SO(16)\times SO(16)$, where the reduction from
$E_8\times E_8$ to $SO(16)\times SO(16)$ is realised in
terms of a Generalised GSO (GGSO) phase in the partition function.
This breaking can be realised in the orbifold formalism
by starting with the partition function of the
$E_8\times E_8$ heterotic string
and acting with a freely acting $Z_2\times
Z_2$ twist given by 
$Z_2^a\times~Z_2^b~:~= 
(-1)^{F_{\xi^1}}\delta ~\times~ (-1)^{F_{\xi^2}}\delta \,$
with 
$\delta X_9 = X_9 +\pi R_9~,$
where $F_{\xi^{1,2}}$ are fermion numbers acting on the first and
second $E_8$ factors,
and $\delta$ is a mod 2 shift in one internal direction 
\cite{parti}.
The action of the $Z_2\times Z_2$
freely acting twist reduces the $E_8\times E_8$ symmetry to
$SO(16)\times SO(16)$. 
The next step in the construction employs a $Z_2^c$ twist
that acts on the internal 
coordinates and produces 16 fixed points. The observable $E_8$, or 
$SO(16)$, gauge symmetries are broken to 
$E_7\times SU(2)$, or $SO(12)\times SO(4)$, respectively.
The matter and Higgs states in the first case are in the 56
representation of $E_7$,
which breaks as $56 = (32,1) + (12,2)$ under $SO(12)\times SU(2)$, where
the $32$ spinorial, and the $12$ vectorial, representations of $SO(12)$
contain the matter and Higgs representations of the Standard Model,
respectively. 
Hence, the freely acting $Z_2^a\times Z_2^b$ twists induce a string
Higgs--matter
splitting mechanism, similar to the string doublet--triplet splitting
mechanism \cite{ps}. 
It turns out that simply adding the $Z_2^c$ twist, {\it i.e.} taking
$[Z_+/(Z_2^a\times Z_2^b)]/Z_2^c$, where $Z_+$ is the $E_8\times E_8$ 
partition function, projects the twisted massless spinorial representations
and keeps the vectorials. 
To restore the spinorial matter representations one needs to analyse
the full $Z_+/(Z_2^a\times Z_2^b\times Z_2^c)$ partition function shown
in figure \ref{fig1}
\begin{vchfigure}
\includegraphics[width=.6\textwidth, height=8cm]{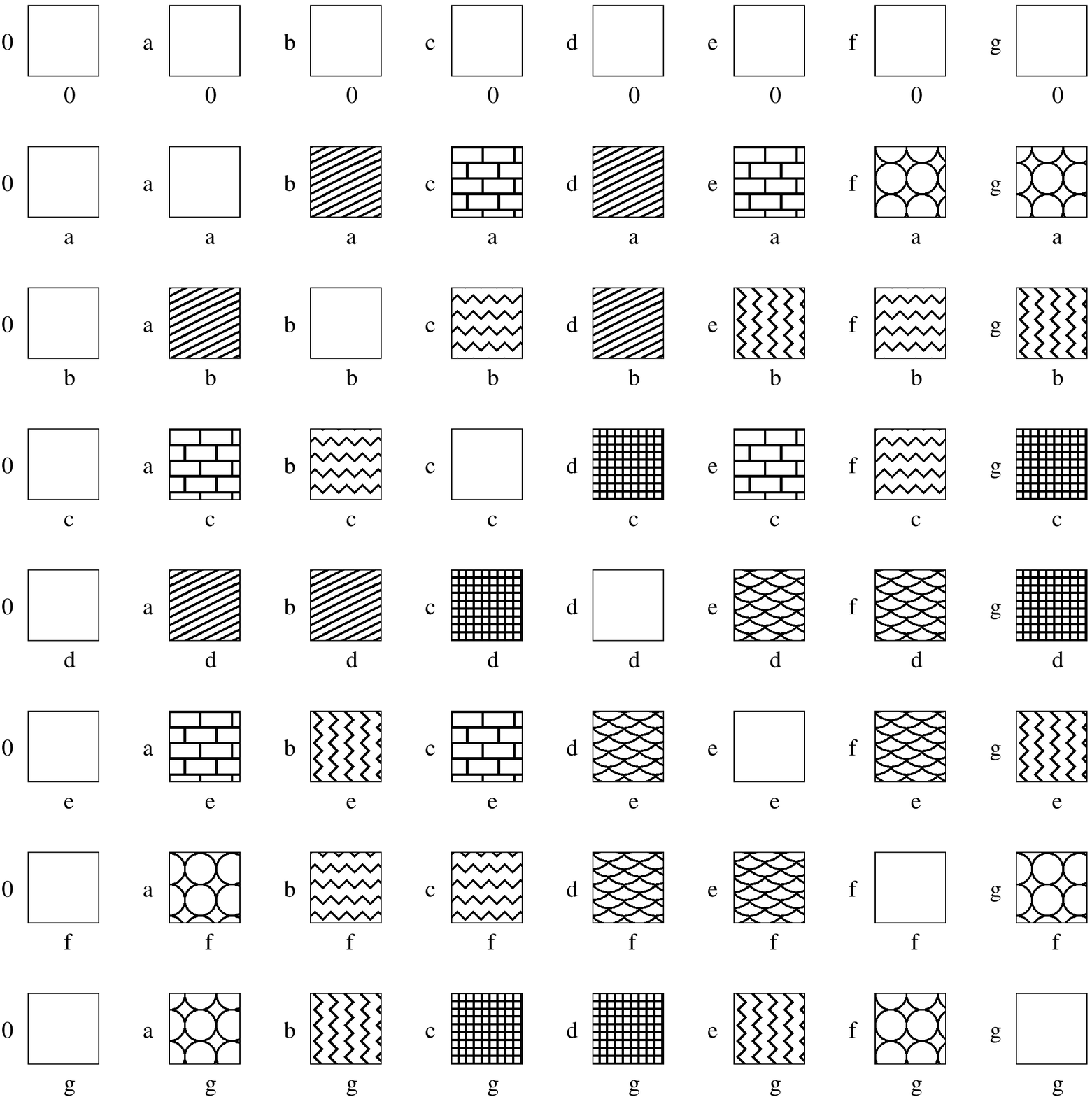}
\caption{The full $Z_2^a\times Z_2^b\times Z_2^c$ partition function}
\label{fig1}
\end{vchfigure}
and take into account 
the eight independent modular orbits (seven discrete torsions) \cite{aft}.
The partition function gives rise to massless states arising from the terms 
{\begin{eqnarray}
{
Q_c 
\left({1\over2}\left(    \left\vert{{2\eta}\over\theta_4}\right\vert^4
                      +
                         \left\vert{{2\eta}\over\theta_3}\right\vert^4\right)
    \Lambda_{p,q}
                \left(
                          {{\epsilon_{06}^- 
                      + 
                            \epsilon_{24}^+(-1)^m}\over2}
                \right)\Lambda_{m,n} 
               \right)
                    {\overline V}_{12}{\overline S}_4{\overline O}_{16} }
& &\label{voo}\\ 
{
Q_c 
\left({1\over2}\left(    \left\vert{{2\eta}\over\theta_4}\right\vert^4
                      +
                         \left\vert{{2\eta}\over\theta_3}\right\vert^4\right)
    \Lambda_{p,q}
                \left(
                          {{\epsilon_{06}^+ 
                      -
                            \epsilon_{24}^-(-1)^m}\over2}
                \right)\Lambda_{m,n} 
               \right)
                    {\overline S}_{12}{\overline O}_4{\overline O}_{16} }
& & \label{soo}
\end{eqnarray}
}
where $Q_c$ produces a $N=2$ supersymmetric space--time spinor; 
$\Lambda_{r,s}$ are $S^1$ lattice sums over momentum and winding
modes, and the $\{O,V,S,C\}$ are the
$\{$singlet, vector, spinor, anti--spinor$\}$ level--one Kac--Moody
$SO(2n)$ characters. The $\epsilon_{ij}^\pm$ are given in terms of
$\epsilon_0=+1$, and the seven
discrete torsions $\epsilon_i=\pm1$, $i=1, \cdots, 7$, as 
$$\epsilon_{ij}^\pm={{\epsilon_i\pm\epsilon_j}\over 2}.$$
It is noted from the partition function and
eqs. (\ref{voo}) and (\ref{soo}) that the case with 
$\epsilon_1=+1,~
\epsilon_2=-1,~
\epsilon_3=+1,~
\epsilon_4=+1,~
\epsilon_5=-1,~
\epsilon_6=+1,~
\epsilon_7=-1,$ 
produces massless spinors and massive vectors, whereas the choice
$\epsilon_1=+1,~
\epsilon_2=+1,~
\epsilon_3=+1,~
\epsilon_4=+1,~
\epsilon_5=+1,~
\epsilon_6=-1,~
\epsilon_7=+1,~$
produces massive spinors and massless vectors. This result arises
because in the first case the massless lattice modes attach to 
the spinor character, whereas the vector character is attached
to massive modes only, and vica versa in the second case. Hence, 
in terms of the discrete torsions the spinor--vector duality map
is realised by the discrete transformation
$
\{ \epsilon_2, \epsilon_5, \epsilon_6, \epsilon_7\}
\rightarrow 
-\{ \epsilon_2, \epsilon_5, \epsilon_6, \epsilon_7\}
$. Additional constraints on
the remaining discrete torsions are imposed by the requirement
that the gauge symmetry is not enhanced, and by space--time spin--statistics.

Counting the number of states in each of the cases in eqs. (\ref{voo})
and (\ref{soo}) we note that there is a mismatch between the two cases.
The spinor and vector representations of $SO(12)$ contain 32 and 12 
states, respectively. The states arising in eq. (\ref{voo}) transform 
as a spinor of the $SO(4)$ group and therefore the total number of states
in this case is 24. There is still a mismatch of eight states between the 
two case. This mismatch is rectified by another term in the partition function
$$
{
Q_c\left( 
 ~{1\over 2}
\left(\left\vert{{2\eta}\over\theta_4}\right\vert^4-
                  \left\vert{{2\eta}\over\theta_3}\right\vert^4
                                        \right)~~\Lambda_{p,q} 
\left({{\epsilon_{06}^-+\epsilon_{24}^+(-1)^m}\over {2~~~~~~~~}}\right)\Lambda_{m,n}
             {\overline O}_{12}{\overline S}_{4}{\overline O}_{16}\right)}.
$$
The first excited twisted lattice modes that arise from this term 
produce eight massless $SO(12)\times SO(16)$ singlets. 
We find that the total number of states is 32 in the two cases.
Two additional solutions \cite{aft} that produce vectorial multiplets of
the hidden $SO(16)$ group, and transform as $SO(4)$ spinors, also give
rise to a total of 32 states. Hence, in all these cases the total number
of states is identical. 

While in this note I discussed merely one illustrative example, 
the free fermionic classification accesses a much larger space
of vacua \cite{classification, spduality}.
A similar spinor--vector duality symmetry is observed
in the larger space of solutions under exchange of free
fermion phases. While the corresponding effective field 
theories are entirely distinct, the string vacua are 
connected via the duality transformations. The full string
theory can access and exchange massless and massive modes
that are not seen in the effective low energy field theory.
Furthermore, the number of states in the dual vacua
is preserved under the duality map.
Thus, while two
vacua may seem entirely distinct from the point of 
view of the low energy physics, they are in fact
equivalent from the string theory point of view.
Further elaboration on this observation is given in ref \cite{ffmt}.
We may surmise 
that from the string point of view the organisation of the states
into multiplets of the underlying gauge symmetry is secondary, whereas
the string internal consistency, which amounts to preservation of the 
number of degrees of freedom under the map is primary. 

\begin{acknowledgement}
Supported by the 
STFC grant \# ST/G00062X/1 and the EU
contract \# MRTN-CT-2006-035863. 

\end{acknowledgement}

\end{document}